# Denoising diffusion-based MRI to CT image translation enables automated spinal segmentation.

- Robert Graf M. Sc* (1)
- Joachim Schmitt (1)
- Dr. Sarah Schlaeger (1)
- Hendrik Kristian Möller M. Sc (1)
- Vasiliki Sideri-Lampretsa M. Sc (4)
- Anjany Sekuboyina M. Sc (1, 3)
- Prof. Dr. Sandro Manuel Krieg (2)
- PD Dr. Benedikt Wiestler (1)
- Prof. Dr. Bjoern Menze (3)
- Prof. Dr. Daniel Rueckert (4, 5)
- Prof. Dr. Jan Stefan Kirschke (1)

(1) Department of Diagnostic and Interventional Neuroradiology, School of Medicine, Technical University of Munich, Germany
(2) Department of Neurosurgery, Klinikum rechts der Isar, School of Medicine, Technical University of Munich, Germany
(3) Department of Quantitative Biomedicine, University of Zurich, Switzerland
(4) Institut für KI und Informatik in der Medizin, Klinikum rechts der Isar, Technical University of Munich, Germany
(5) Professor of Visual Information Processing, Imperial College London

## Abstract

**Background:** Automated segmentation of spinal MR images plays a vital role both scientifically and clinically. However, accurately delineating posterior spine structures presents challenges.

**Methods:** This retrospective study, approved by the ethical committee, involved translating T1w and T2w MR image series into CT images in a total of n=263 pairs of CT/MR series. Landmark-based registration was performed to align image pairs. We compared 2D paired (Pix2Pix, denoising diffusion implicit models (DDIM) image mode, DDIM noise mode) and unpaired (contrastive unpaired translation, SynDiff) image-to-image translation using "peak signal to noise ratio" (PSNR) as quality measure. A publicly available segmentation network segmented the synthesized CT datasets, and Dice scores were evaluated on in-house test sets and the "MRSpineSeg Challenge" volumes. The 2D findings were extended to 3D Pix2Pix and DDIM.

**Results:** 2D paired methods and SynDiff exhibited similar translation performance and Dice scores on paired data. DDIM image mode achieved the highest image quality. SynDiff, Pix2Pix, and DDIM image mode demonstrated similar Dice scores (0.77). For craniocaudal axis rotations, at least two landmarks per vertebra were required for registration. The 3D translation outperformed the 2D approach, resulting in improved Dice scores (0.80) and anatomically accurate segmentations in a higher resolution than the original MR image.

**Conclusion:** Two landmarks per vertebra registration enabled paired image-to-image translation from MR to CT and outperformed all unpaired approaches. The 3D techniques provided anatomically correct segmentations, avoiding underprediction of small structures like the spinous process.

**Keywords:**

Image-to-image, spine segmentation, MRI to CT, point-registration, denoising diffusion.

**Key Points:**

- Unpaired image translation lacks in converting spine MRI to CT effectively.
- Paired translation needs registration with two landmarks per vertebra at least.
- Paired image-to-image enables segmentation transfer to other domains.
- 3D translation enables super resolution from MRI to CT
- 3D translation prevents under-prediction of small structures.

**Abbreviations:**

| **DDIM** | denoising diffusion implicit model |
|---|---|
| **CUT** | contrastive unpaired translation |
| **Pix2Pix** | (Proper name) |
| **SynDiff** | (Proper name) |
| **SA-UNet** | self-attention U-network |
| **MRSSegClg** | MRSpineSeg Challenge |
| **PSNR** | peak signal-to-noise ratio |
| **CM** | center of mass |
| **GNC** | German National Cohort |
| **GAN** | Generative Adversarial Network |

**Data available:**

The MRSSegClg dataset is available under: www.spinesegmentation-challenge.com/ [24]

The used segmentation algorithm can be accessed by: https://anduin.bonescreen.de/ [3, 4].

Our code for registration and our deep learning methods are available under: Point registration; URL: https://github.com/robert-graf/Pointregistation DOI: 10.5281/zenodo.8198697 - Platform independent; Python 3.10 or higher with packages simpleitk nibabel jupyter simpleitk pillow pyparsing matplotlib; License: MIT License

Readable Conditional Denoising Diffusion – URL https://github.com/robert-graf/Readable-Conditional-Denoising-Diffusion – https://doi.org/10.5281/zenodo.8221159 - Platform independent – Python 3.10 or higher with packages pytorch pytorch-lightning numpy configargparse einops ipykernel ipython joblib nibabel pandas scikit-image scikit-learn scipy tqdm ema-pytorch; License: MIT License

Other used publicly available algorithms:

SynDiff: https://github.com/icon-lab/SynDiff (17) Platform independent – Python>=3.6.9 with packages torch>=1.7.1 torchvision>=0.8.2 cuda=>11.2 ninja

Deformable data argumentation: https://pypi.org/project/elasticdeform/ (26) DOI: 10.5281/zenodo.4563333 Platform independent – Python package

# Background

The different contrasts of CT and MRI offer distinct clinical utilities. Generally, segmentation is a prerequisite to automatically extract biomarkers, especially in large cohorts like the German National Cohort (GNC)[1] or the UK-biobank[2]. While the extraction of the precise bone structure of the spine from CT is publicly available [3, 4], neither a segmentation nor an annotated ground truth dataset for the whole spine including the posterior elements is currently available for MRI. Accurate segmentations are not only vital for scientific studies but also enable the exact localization of abnormalities in clinical routine. Unlike CT, MRI provides additional information about bone marrow edema-like changes, intervertebral disc degeneration, degenerative endplate changes, ligaments, joint effusions, and the spinal cord. Robust and precise segmentation and quantification of such spinal structures is a prerequisite e.g. to evaluate large epidemiologic studies or to enable automated reporting. An alternative to labor-intensive manual annotations is the potential use of image-to-image translation to extract bony structures. This approach may overcome challenges like partial volume effects (e.g. at the spinous process) and subtle signal differences (e.g. of vertebral endplates and ligaments in MRI), which are easily distinguishable in high-resolution CT but not in MRI.

Image-to-image translation involves transforming images from one domain to another, and several deep learning methods have been employed for this purpose, including Pix2Pix[5], CycleGAN[6], and contrastive unpaired translation (CUT) [7]. These methods have been used in various studies to generate missing sequences, translate to different domains, enhance image quality, and improve resolution[8]. In the medical domain, these methods have shown success in rigid structures like the brain, head, and pelvis, where registration guarantees that both domains have similar tissue distributions and anomalies [8]. However, if biases are not accounted for, the model may hallucinate new structures to fit both distributions [9]. Due to this difficulty, translating warpable structures like the spine is less explored in the literature. Some successful implementations have shown that translated images can be similar to the target images and might mislead medical experts [10–14]. However, none of these works have focused on using translations for downstream tasks, such as segmentations in the output domain.

This study aimed to develop and compare different image translation networks for pretrained CT-based segmentation models when applied to MRI datasets (Figure 1). The primary focus was on segmenting the entire spine, with special attention to accurately translating the posterior spine structures, as they pose challenges in MRI delineation. We compared GAN-based approaches [5, 7] with new denoising diffusion models [15–17]. Denoising diffusion functions are fundamentally different from GANs, as they add and remove noise to an image instead of relying on the discriminator and generator zero-sum game in GANs. In the computer vision domain, denoising diffusion models have outperformed GANs in various tasks, including upscaling, inpainting, image restoration, and paired image-to-image translation [18]. While diffusion has been applied to medical image translation tasks in a limited number of papers [17, 19–22], we adapted the conditional denoising diffusion for paired image-to-image translation in 2D and 3D.

The purpose of this paper was (1) to improve existing image-to-image translation for spine MRI to CT translation by improving all steps of the process: from data alignment, implementation of new denoising diffusion translations and comparison to GANs, and finally extension of our findings to 3D translation. (2) To utilize the translated CT images for automatic segmentation of the entire spine, eliminating the need for a manually labeled segmentation mask in the original MRI domain. (3) To develop the ability to generate full spine segmentations on MRI, which are currently not available.

# Materials and Methods

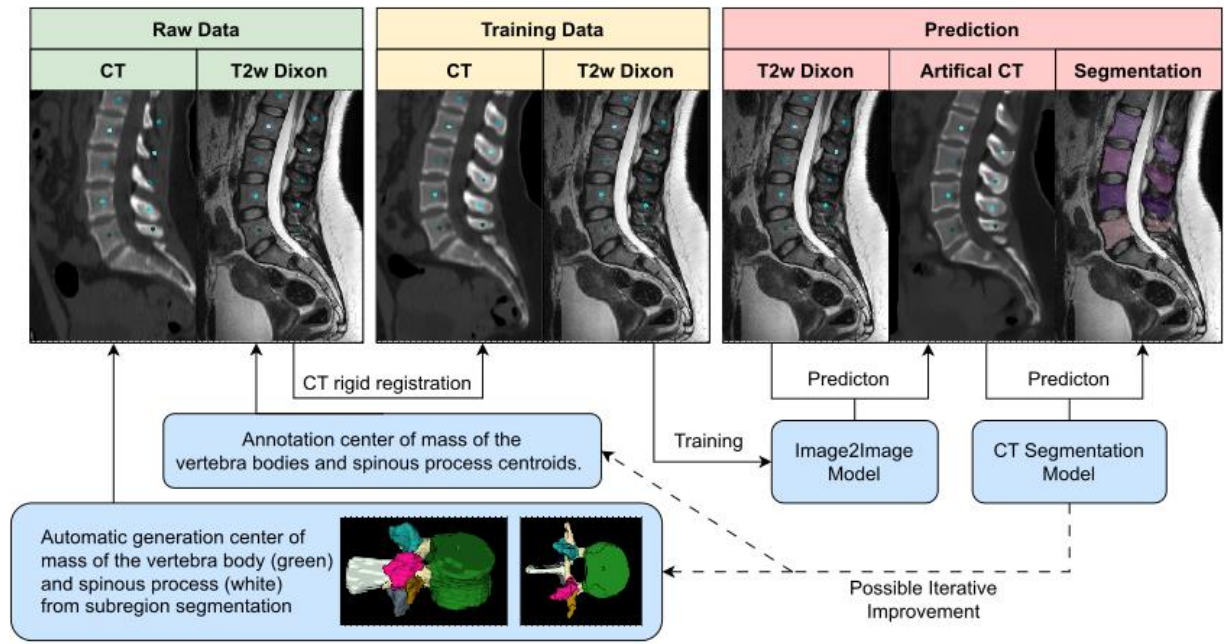

**Figure 1:** Our training pipeline. *In our datasets, we identified the center of the vertebral body and spinous process (green box; raw data). Based on the center points, we rigidly registered CT onto MRI to align the bone structures between the two images (yellow box; training data). Aligned images were used to train our image-to-image models. Finally, the MRIs of validation and test sets were translated to CT images. Segmentation was performed on synthesized CT images, and consequently, were perfectly aligned with the original MRIs (blue box from left to right; prediction). The generated segmentations can be used for generating additional and new center points to iteratively optimize the registration.*

In brief, we aligned CT and MR spine images through rigid landmark registration[23]. With this paired data, we trained various image-to-image models to generate CT images. We used an available CT segmentation algorithm [3, 4] to generate vertebral masks in these synthesized CTs for the original MRI. These resulting segmentations were subsequently used to generate new landmarks for new training data (Figure 1). We compared different landmark registrations and 2D models. Finally, we adapted the results into 3D models and assessed the accuracy of the resulting segmentations.

## Data

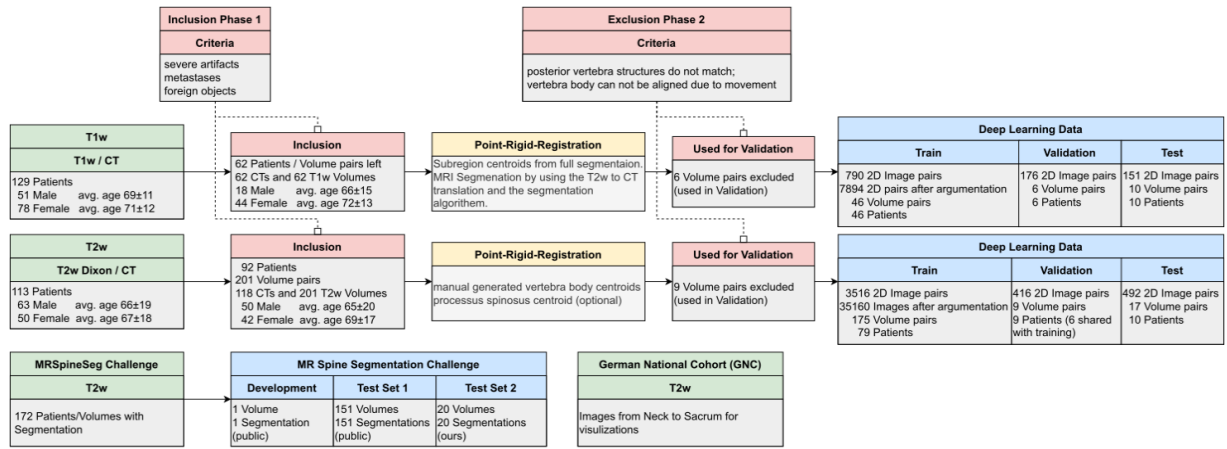

**Figure 2:** Datasets, preparation, exclusion, and split. *MR data were acquired with 12 different scanners from 3 different vendors. Additionally, we used the MRSSegClg for external testing. For the 2D training we only consider 2D slices containing a spine. We demonstrated generalizability using a full-body MRI from the German National Cohort (GNC) dataset for the figures in this paper.*

In this study, we retrospectively collected sagittal T1w/T2w MR and corresponding CT images of the spine from the same patient within a week. Approval from the local ethics committee was obtained, and informed consent was waived. Figure 2 illustrates our data selection process. 62 T1w image series (18 males and 44 females; average age 66±15y / 72±13y) were used from another unpublished in-house study, including five thoracic and 57 lumbar volumes. Additionally, a new dataset was collected of 201 T2w image series (50 males and 42 females; average age 65±20y / 69±17y) from 92 patients, including 38 cervical, 99 thoracic, and 70 lumbar volumes. Patients with fractures and degenerative changes were included, while those with motion artifacts, metastases, and foreign objects were excluded, because for segmentation models, it would benefit when the translation suppresses these anomalies. We performed rigid registration of the matching MRIs and CTs based on the center of mass of the vertebral body and the spinous process. (See Figure 1 bottom left). In-house test set, training, and validation set were split patient-wise for different MR acquisitions of other spine regions. For validation, six T1w and nine T2w MRIs were used as they could not be aligned with the CTs due to substantially different patient positioning.

We used 172 lumbar MR and segmentation volumes from the MRSpineSeg Challenge (MRSSegClg) [24, 25] for external evaluation of Dice-Scores. This dataset focuses on the lumbar region, but the segmentation exceeds the bony borders, questioning its validity. One subject was used for pipeline development and validation. Validation sets were used to find optimal inference parameters and to avoid overfitting. Since the labels in MRSSegClg encompass not only the bony spine but also adjacent ligaments and soft tissue, we manually adjusted the labels for a subset of 20 volumes to restrict them solely to the bone. We analyzed these subsets as two distinct datasets.

**Image preprocessing**

CT and MR datasets were rigidly registered[23] by using landmarks to facilitate paired image translation. For the single-landmark approach, we selected the center of mass (CM) of the vertebral bodies. To address rotational misalignment around the cranio-caudal axis, the CM of the spinal processes was added for the two-landmark approach, as such rotational misalignment was frequently observed. Landmarks for CT were automatically determined based on vertebral and subregion segmentations (Figure 1). For the T2w images, we manually identified the CM points for both the vertebral bodies and the spinous processes. The manual centroid selection and ground truth segmentation corrections in the test sets were performed by J. S., a radiologist with three years of experience. To obtain the points for the T1w image, we synthesized them by adapting the T2w to CT translation, generating segmentation, and extracting the CM for T1w. Roughly 10 to 20% of the failure cases were first excluded then translated with models that was trained on the other T1w images. This proved sufficient to generate all CM points. To assess the impact of additional landmarks on registration, we computed the Dice score using our pipeline on the T2w dataset using the manual ground truth as a reference.

CT images were transformed to the range of [-1, 1] by dividing the values by 1000 HU and clamping outliers to retain air, soft tissue, and bone while suppressing extreme intensities. Linear rescaling was applied to the MRI data, converting the range from [max, 0] to [-1, 1]. To account for varying intensities, MRIs were augmented with a random color jitter (brightness, contrast randomization: 0.2). Image pairs were resampled to a uniform spatial resolution of 1x1 millimeters in the sagittal plane and a slice thickness of 2.5-3.5 millimeters, as acquired in the MRI. To enhance the training data by a factor of 10 and simulate weak scoliosis and unaligned acquisition, we introduced 3D image deformations using the elastic deformation Python plugin[26]. Subsequently, the volumes were sliced into 2D sagittal images, and slices without segmentation were removed. Random cropping was performed to adjust the image size to 256x256 pixels.

## Models for Image-to-image Translation

To compare various image-to-image translation methods, we implemented two unpaired methods, namely CUT[7] and SynDiff[17], along with three paired methods, Pix2Pix[5], DDIM noise, and DDIM image. The training process involved unregistered and registered data using both single- and two-landmark approaches. For DDIM, we employed a UNet[26] architecture with convolutional self-attention and embeddings for the timesteps, which we refer to as self-attention U-network (SA-UNet) [18, 27, 28]. The Diffusion mechanism predicted either noise or the image, with the other computed during inference. A learning rate of 0.00002 was used, and we set the timestep to t=20 for the DDIM inference parameter. The value of $\eta = 1$ (noise generation is fully random) was determined by optimizing on the validation set. We compared our approach to CUT[7], Pix2Pix[5], and SynDiff[17]. During our experiments, we performed a hyperparameter search for the reference ResNet and UNet. Additionally, we introduced a weighted structural similarity index metric (SSIM) loss from a recent paper[29] to update the loss formulation. To further explore the impact of different models and methods, we also tested CUT and Pix2Pix with the SA-UNet. All models were randomly initialized. In our analysis of DDIM, we ablated three inference parameters[16, 30]. However, the results did not show substantial effects, and we have included them in the supplementary material along with brief descriptions of the tested methods.

**Image Quality**

The evaluation of image quality involved comparing actual and synthesized CT images. To quantify this, we used the "peak signal-to-noise ratio" (PSNR) metric. In this context, the reference image serves as the "signal," and the divergence between the two images is considered the "noise." A PSNR value above 30 dB indicates that the difference between the two images is imperceptible to the human eye [10]. It's important to note that we did not control the correspondence of soft tissue, as it fell outside the scope of our downstream task. To handle this in our evaluation, we masked pixels that were further than 10 pixels away from a segmented spine structure, setting them to zero. We also computed the absolute difference (L1) mean squared error (MSE) structural similarity index measure (SSIM) and Visual Information Fidelity (VIFp).

**Downstream task: Segmentation**

We utilized a publicly available segmentation algorithm[3, 4] on the synthesized CT images. We then compared the Dice scores globally and on a vertebral level between the synthesized and ground truth segmentations in four datasets. (1,2) The segmentation ground truth of the in-house datasets was derived from the aligned CT image and was manually corrected. (3) The segmentation of the MRSSegClg that is known to exceed the bony structures and (4) a manually corrected subset of MRSSegClg [24, 25]. In Figure 3 C/D the segmentation reaching beyond the bony structures of MRSSegClg is highlighted. For analysis purposes, we excluded structures that the CT segmentation algorithm could not segment, such as the sacrum and partially visualized vertebrae.

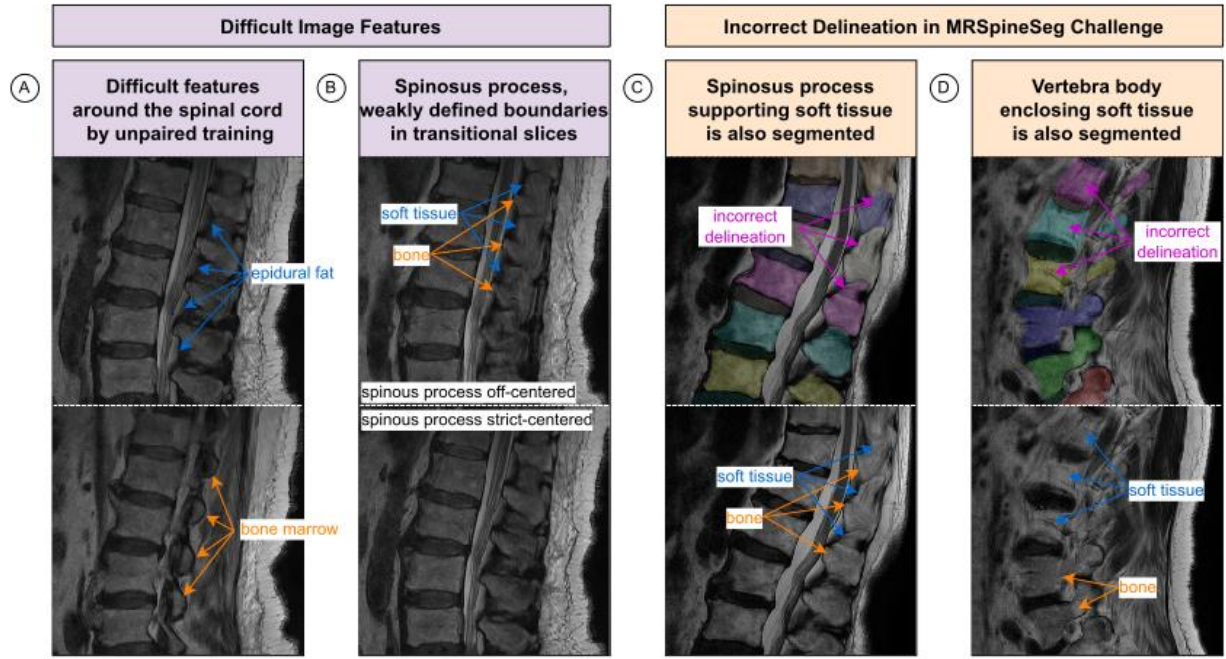

**Figure 3:** Difficulties of the MRI data for unpaired training and issues with the MRSSegClg segmentation. *A) The bone marrow of the posterior elements and the epidural fat were not easily differentiated. Unpaired learning has issues translating the arcus as bone and the epidural fat as soft tissue in the CT domain. B) In posterior elements, bone and soft tissue boundaries are weakly defined due to partial volume effects in and around the spinous process. C) The segmentations of the MRSSegClg include soft tissues around the spinous process, caused by difficulties of the original annotators as described in B. D) The soft tissues around the vertebrae are also segmented in the MRSSegClg. C and D are the reasons why we manually improved the segmentation in a small subset.*

## 3D Image Translation with Diffusion

The first implementations of both DDIM and Pix2Pix in 3D, similar to the 2D approach, did not converge. We thus implemented changes according to recommendations of Bieder et al. [31]. To optimize GPU storage, we eliminated attention layers and replaced concatenation skip connections with addition operations. Additionally, we introduced a position embedding by concatenating ramps ranging from zero to one of the original images' full dimensions into the input. The training was done on 3D patches and our approach used a patch size of (128x128x32), where the left/right side was limited to 32 pixels due to the image shape. This setup is "fully convolutional", which means that during inference, an image of any size can be computed by the network as long the sides are divisible by 8. To the best of our knowledge, this represents the first 3D image-to-image translation with diffusion. Since 3D translations require to include the left/right direction, we resampled all images to 1 mm isotropic.

## Statistical Analysis and Software

We employed a paired t-test to assess the significance of PSNR and Dice score between different models. To achieve a fixed size of 256x256 pixels for assessing image quality, we used one crop per image slice. When reporting differences in multiple experiments, we present the worst (i.e., highest) p-value. We skip significance calculations other image quality metrics because the results are redundant. For 3D data, we pad the test data, and the 3D models generate 1 mm isotropic volumes, which are later resampled to the original MRI size.

# Results

## Influence of Rigid Registration

Networks trained on unregistered data were incapable of learning the difference between soft tissue and bone. During our early testing, we noticed that most methods could correctly identify the vertebral body, but translating the posterior structures was impossible. Especially the spinous process was often omitted in the translation, as shown in Figure 4. "One point per vertebra" registration was sufficient for the vertebral body translation, but the spine could rotate around the cranio-caudal axis. This caused the spinous process to disappear in translated images (see Figure 4 A/B). Additionally, confusion between epidural fat and bone shifted the entire posterior elements towards the spinal cord. Overcoming this issue required accounting for rotation by adding additional points to the rigid registration (Figure 4). Next to visual findings, we observed a significant increase in Dice from one to two points per vertebra registration. Pix2Pix 0.68 to 0.73 ($p<0.003$); SynDiff 0.74 to 0.77 ($p<0.001$); DDIM noise 0.55 to 0.72 ($p<0.011$) and DDIM image 0.70 to 0.75 ($p<0.001$). Notably, the best unpaired method, SynDiff, could not learn posterior structure translation without registration. (Dice without registration 0.75).

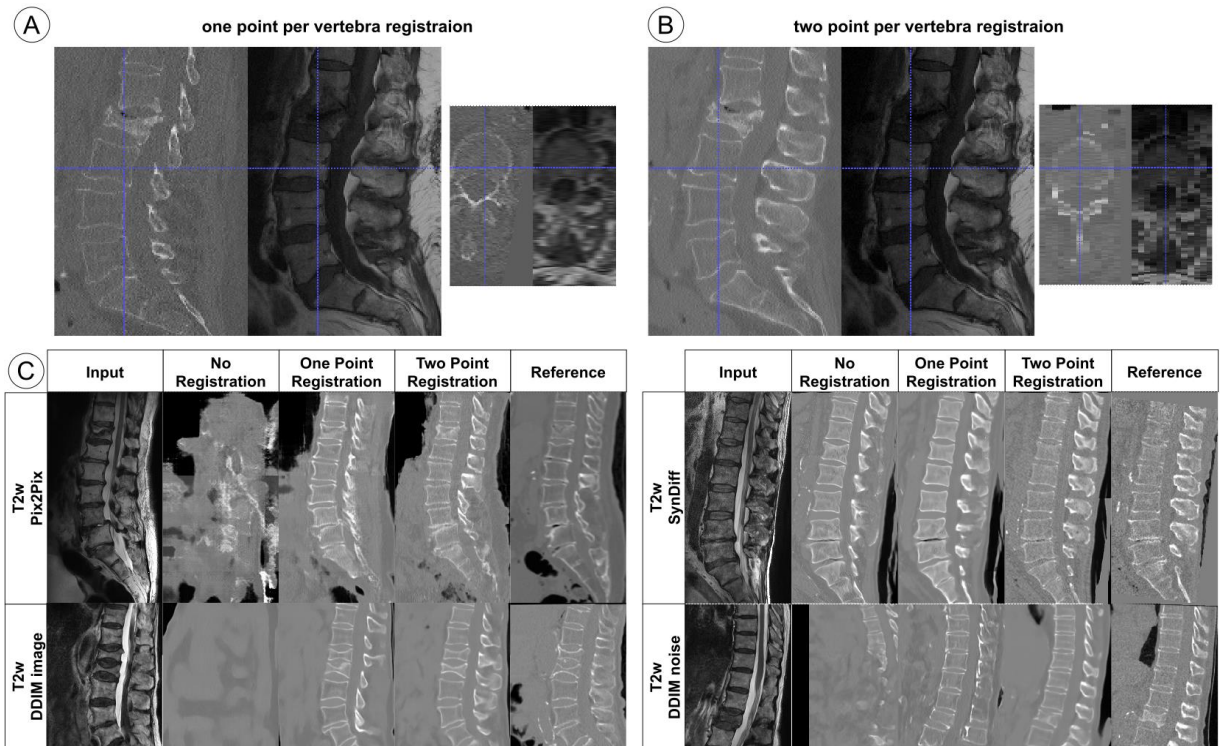

**Figure 4:** Comparison of one and two registration points per vertebra on real data. ***A:*** *We registered with a single point in the center of the vertebral body. The vertebral body could rotate along the spine axis. This caused the posterior vertebra structures to be misaligned.* ***B:*** *When we registered the images with an additional point on the spinous process, we avoided this rotation around the spine itself. The blue dashed lines are for locating the relation between axial and sagittal slices.* ***C:*** *Translation with networks trained on registrations with zero, one, or two points per vertebra. Images are from the in-house T2w test dataset. Posterior structures are only reconstructed correctly with two-point registration. DDIM = denoising diffusion implicit model*

**Image Quality**

The unpaired CUT models performed worse than all others (p<0.001), while all other models performed on a similar level. See Table 1 for PSNR and other common metrics. Example outputs from the test sets can be seen in Figure 5. The Pix2Pix with the SA-UNet performed better on T1 and worse on T2 than the smaller UNet (T1: p<0.001; T2: p=0.041). Even though SynDiff had an unpaired formulation, it had similar results compared to our paired

Pix2Pix, DDIM noise (slightly worse in T1w and better in T2w, all p<0.003). The DDIM image mode performed slightly better than the DDIM noise mode (p<0.001), SynDiff (p<0.001), and Pix2Pix (p<0.001). DDIM image mode produces images with less noise than the original data. Less noise should make the segmentation easier. Overall, the DDIM image mode was our best-performing 2D model.

**Table 1:** Image Quality for T1w and T2w to CT translation.

| From T1w | L1↓ | MSE ↓ | PSNR↑ | SSIM↑ | VIFp ↑ |
|---|---|---|---|---|---|
| CUT ResNN (unpaired) | 0.0224 | 0.0050 | 23.50 | 0.835 | 0.295 |
| CUT SA-UNet (unpaired) | 0.0295 | 0.0083 | 21.76 | 0.819 | 0.269 |
| Pix2Pix UNet | 0.0143 | 0.0023 | 27.37 | 0.881 | 0.392 |
| Pix2Pix SA-UNet | 0.0135 | ∴**0.0020** | 27.82 | 0.883 | 0.394 |
| SynDiff (unpaired) | 0.0150 | 0.0024 | 27.01 | 0.865 | 0.373 |
| DDIM noise $\eta=1$, t=20, w=0 | 0.0136 | 0.0021 | 27.60 | 0.879 | 0.396 |
| DDIM image $\eta=1$, t=20, w=0 | ∴**0.0131** | ∴**0.0020** | ∴**27.89** | ∴**0.887** | ∴**0.411** |

| From T2w | L1↓ | MSE ↓ | PSNR↑ | SSIM↑ | VIFp ↑ |
|---|---|---|---|---|---|
| CUT ResNN (unpaired) | 0.0213 | 0.0046 | 23.72 | 0.848 | 0.312 |
| CUT SA-UNet (unpaired) | 0.0215 | 0.0046 | 23.75 | 0.850 | 0.311 |
| Pix2Pix UNet | 0.0142 | 0.0023 | 26.95 | 0.895 | 0.392 |
| Pix2Pix SA-UNet | 0.0142 | 0.0023 | 26.87 | 0.890 | 0.384 |
| SynDiff (unpaired) | 0.0140 | 0.0022 | 27.12 | 0.885 | 0.385 |
| DDIM noise $\eta=1$, t=20, w=0 | 0.0139 | 0.0023 | 26.92 | 0.894 | 0.391 |
| DDIM image $\eta=1$, t=20, w=0 | ∴**0.0131** | ∴**0.0021** | ∴**27.36** | ∴**0.898** | 0.401 |
| | | | | | |
| Pix2Pix 3D | 0.0188 | 0.0039 | 26,38 | 0.889 | 0.428 |
| DDIM 3D noise $\eta=1$, t=25 | 0.0194 | 0.0041 | 26,22 | 0.894 | ∴**0.444** |
| DDIM 3D image $\eta=1$, t=25 | 0.0189 | 0.0040 | 26,22 | 0.892 | 0.434 |

*Arrows indicate if smaller or bigger is better. As a visual aid, we marked the best values with ∴. We marked multiple values if they were below the rounding threshold. The ground truth is registered real CTs. The image pairs are from the test set of our in-house data. MSE = mean squared error, PSNR = peak signal-to-noise ratio, SSIM = structural similarity index metric, VIFp = visual information fidelity, DDIM = denoising diffusion implicit model, DDPM = denoising diffusion probabilistic model, CUT = contrastive unpaired translation, SA-UNet = self-attention U-network*

**Figure 5:** Translation from test sets T1w/T2w to CT from the neck to the lumbar vertebra. *We did not control the type of reconstruction of the CT. Therefore, the noise level and appearance could differ from the reference and were still considered correct. The 3D variances were trained on an improved training set, which was only done for T2w. The reference is a registered real CT. * is an off-angle acquisition with strong partial volume effects. The data set contains a high number of broken vertebral bodies, with causes them to be also translated correctly. DDIM = denoising diffusion implicit model, CUT = contrastive unpaired translation, SA-UNet = self-attention U-network*

**Downstream task: Segmentation**

Three 2D models shared the best Dice score: Pix2Pix SA-UNet, SynDiff, and DDIM image mode. See Table 2; (Pix2Pix SA-UNet vs. SynDiff p=0.019; Pix2Pix SA-UNet vs. DDIM image mode p<0.001; DDIM image mode vs. SynDiff p=0.455). DDIM in noise mode and Pix2Pix UNet (DDIM noise vs. Pix2Pix UNet p=0.972) were worse than the three best models (p<0.001). The CUT reconstruction was unsuited for segmentation and was the worst model (CUT vs. all p<0.001). An example of the segmentation from different translations for a full spine can be found in Figure 6 in an example dataset from the GNC[1].

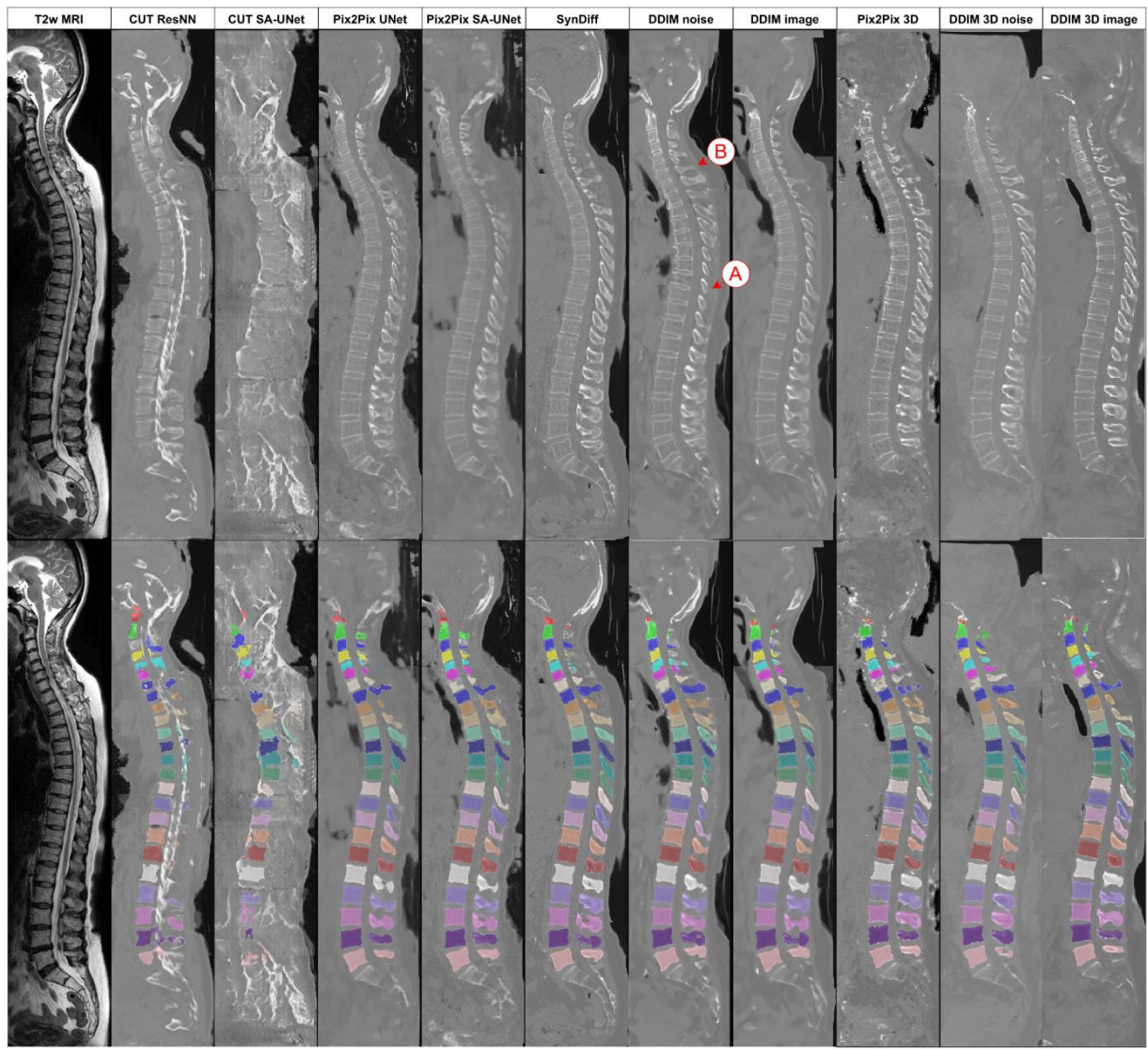

**Figure 6:** Translation from T2w MR to CT and the segmentation results in an external full spine scan. *The MRI shown is a random image from the GNC dataset. The CT translation is stitched. The 2D networks only work on a fixed size of 256x256, and the 3D models run out of memory for the entire image. The 2D networks needed classifier free-guidance (w=1) for these out-of-distribution images or else the neck regions would not form correctly because the frontal area has a drop in MR signal. The 3D networks don't delineate the background and soft tissue when we use a small number of steps (t=25).* **A:** *We observed underpredictions in the thorax process spinous* **B:** *The neck has higher variability between different translations. Moving to 3D translation resolves these issues. DDIM = denoising diffusion implicit model, CUT = contrastive unpaired translation, SA-UNet = self-attention U-network*

We observed comparable rankings in the MRSSegClg[24, 25] and T1w datasets when excluding the vertebral body (Table 3). In the in-house T2w test set, SynDiff has a considerably higher Dice score than Pix2Pix SA-UNet and DDIM image mode (p<0.001), indicating a better performance in the "more complicated" anatomical structures for this data set only.

The correction of the MRSSegClg segmentations resulted in an increased Dice score of up to 0.02. The rankings of all methods on the original versus the corrected MRSSegClg dataset were mostly consistent, indicating that no method had exploited the false delineation by overpredicting the segmentation.

Overall, Pix2Pix SA-UNet, DDIM image mode, and SynDiff were equally capable of producing CT images for the segmentation algorithm. Closely followed by DDIM noise mode and the Pix2Pix UNet.

## 3D Image Translation with Diffusion

All 3D models increased the Dice scores compared to our 2D models (p < 0.006). Pix2Pix 3D and DDIM 3D noise performed on a similar level, while DDIM 3D image performances were consistently a bit better close to the rounding threshold (p<0.001). PSNR showed a drop compared to the 2D variants. The 3D models outperform all 2D models on posterior structures (See Figure 7 T2w: p<0.024; MRSSegClg (ours): p<0.005 for DDIM 3D image, p<0.062 DDIM 3D noise; p<0.462 Pix2Pix 3D; Posterior structures are unavailable in the original MRSSegClg). With the rescaling to 1 mm isotropic, we receive a super-resolution of our mask in the thick slice direction that resembles a more realistic 3D shape than the native resolution (Figure 7).

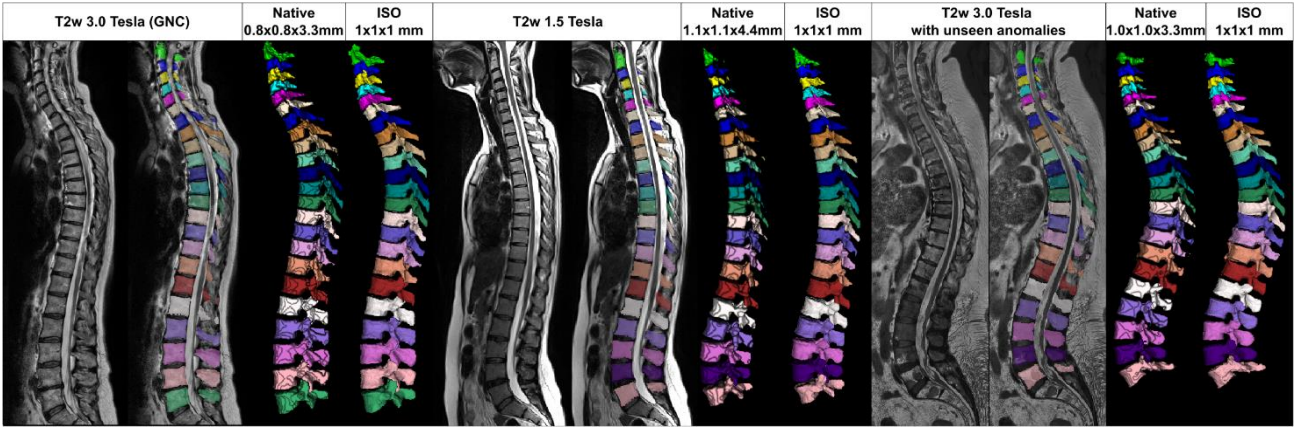

**Figure 7:** 3D visualization of the segmentation from subjects out of the GNC and in-house datasets. *The 3D translation models produce isometric segmentation (iso) that looks biologically correct. After downscaling to the native resolution (native), we observe that the spinous process gets deformed by reducing the slice thickness because the spinous process is thinner than two to three slices. The examples are translated by the DDIM image mode model. We observe no noticeable drop in translation quality for MRIs from other MR scanners. Degenerative changes that are not in the training set are often repaired during translation. While it can partially reproduce when vertebral bodies grow together, which is present in rare cases in the training set. This can be observed by the over-segmentation in the right image from vertebra 7 to 10 counted from the bottom.*

**Table 2**: Average Dice score↑ per volume and per vertebra on the T1w, T2w and the MRSSegClg.

| Dataset | per vol. T1w | per vert. T1w | per vol. T2w | per vert. T2w | per vol. MRSSegClg | per vert. MRSSegClg | per vol. MRSSegClg (our) | per vert. MRSSegClg (our) |
|---|---|---|---|---|---|---|---|---|
| CUT ResNN (unpaired) | 0.30 | 0.28 | 0.49 | 0.46 | 0.54 | 0.49 | 0.54 | 0.50 |
| CUT SA-UNet (unpaired) | 0.09 | 0.08 | 0.26 | 0.23 | 0.02 | 0.01 | 0.03 | 0.02 |
| Pix2Pix UNet | 0.79 | 0.80 | 0.73 | 0.69 | 0.75 | 0.74 | 0.76 | 0.76 |
| Pix2Pix SA-UNet | ∴**0.82** | 0.82 | 0.75 | 0.72 | ∴**0.77** | ∴**0.76** | 0.77 | 0.77 |
| SynDiff (unpaired) | 0.80 | 0.81 | ∴**0.77** | ∴**0.74** | ∴**0.77** | ∴**0.76** | 0.77 | 0.76 |
| DDIM noise η=1, t=20, w=0 | 0.78 | 0.77 | 0.72 | 0.69 | 0.75 | 0.73 | 0.77 | 0.78 |
| DDIM image η=1, t=20, w=0 | ∴**0.82** | ∴**0.83** | 0.75 | 0.72 | ∴**0.77** | ∴**0.76** | ∴**0.78** | 0.78 |

| | | | | | | |
|---|---|---|---|---|---|---|
| Pix2Pix 3D | – | – | 0.79 | 0.78 | 0.78 | **\*0.78** | 0.79 | **\*0.80** |
| DDIM 3D noise η=1, t=25 | – | – | 0.79 | 0.78 | 0.78 | **\*0.78** | **\*0.80** | **\*0.80** |
| DDIM 3D image η=1, t=25 | – | – | **\*0.80** | **\*0.79** | **\*0.79** | **\*0.78** | **\*0.80** | **\*0.80** |

*MRSSegClg (ours) is a split where we improved the segmentation better to align the segmentation with the actual bone structure. T1w and T2w ground truths are corrected segmentations of the registered CTs. We marked the best values for the 2D cases with ∴ and the overall best \*. vol. = volume, vert. = vertebra, MRSSegClg = MRSpineSeg Challenge, DDIM = denoising diffusion implicit model, CUT = contrastive unpaired translation, SA-UNet = self-attention U-network*

**Table 3**: Average posterior structures Dice score↑ per volume and per vertebra

| Dataset | per vol. T1w | per vert. T1w | per vol. T2w | per vert. T2w | per vol. MRSSegClg (our) | per vert. MRSSegClg (our) |
|---|---|---|---|---|---|---|
| CUT ResNN (unpaired) | 0.09 | 0.09 | 0.17 | 0.15 | 0.16 | 0.13 |
| CUT SA-UNet (unpaired) | 0.01 | 0.01 | 0.07 | 0.05 | 0.00 | 0.00 |
| Pix2Pix UNet | 0.64 | 0.62 | 0.55 | 0.50 | 0.56 | 0.55 |
| Pix2Pix SA-UNet | ∴**0.68** | ∴**0.67** | 0.59 | 0.54 | ∴**0.58** | 0.56 |
| SynDiff (unpaired) | 0.67 | ∴**0.67** | ∴**0.63** | ∴**0.58** | ∴**0.58** | ∴**0.57** |
| DDIM noise η=1, t=20, w=0 | 0.61 | 0.59 | 0.50 | 0.46 | 0.57 | 0.56 |
| DDIM image η=1, t=20, w=0 | ∴**0.68** | ∴**0.67** | 0.58 | 0.53 | ∴**0.58** | ∴**0.57** |
| | | | | | | |
| Pix2Pix 3D | – | – | 0.69 | 0.67 | 0.59 | 0.58 |
| DDIM 3D noise η=1, t=25 | – | – | **\*0.70** | **\*0.68** | 0.60 | **\*0.60** |
| DDIM 3D image η=1, t=25 | – | – | **\*0.70** | **\*0.68** | **\*0.61** | **\*0.60** |

*The vertebral body is removed from the calculation by an automatic subregion segmentation on The T1w, T2w, and MRSSegClg (ours). The unchanged MRSSegClg could not be subregion segmented. We marked the best values from 2D cases with ∴ and the overall best \*. vol. = volume, vert. = vertebra, MRSSegClg = MRSpineSeg Challenge, DDIM = denoising diffusion implicit model, CUT = contrastive unpaired translation, SA-UNet = self-attention U-network*

# Discussion

This study successfully demonstrated the feasibility of translating standard sagittal spine MRI into the CT domain, enabling subsequent CT-based image processing. Specifically, the registration process, with a minimum of two points per vertebra, enables accurately translating posterior structures, which are typically challenging for image translation and segmentation. To achieve this, a low-data registration technique was introduced for pairing CT and MRI images, which can be automated by our translation and segmentation pipeline. In our low-data domain, paired translation methods performed on a similar level, with DDIM in image mode being the single best model. The spinous process was not always correctly translated in our 2D approaches. We resolved this issue by changing the process to 3D. Our 3D methods had a drop in image quality compared to the 2D translation. We believe this is due to the required resampling from the 1 mm isotropic output to the native resolution of the test data. Ultimately, the image-to-image translation facilitated MRI segmentation using a pretrained CT segmentation algorithm for all spine regions.

Our results extend prior works that have been limited to high-resolution gradient-echo Dixon T1w sequences to CT translations [14, 32, 33] as well as to intra-modality MR translations for different contrasts from standard T1w and T2w TSE sequences to short tau inversion recovery [34] or T2w fat-saturated images [35], frequently used in spinal MRI. Commercial products are available for MRI to CT translation [36, 37]. However, in contrast to our approach, they require a dedicated, isotropic gradient-echo sequence. They are unavailable for standard T1w or even T2w TSE sequences. Acquiring an additional, dedicated image only for segmentation is resource and time demanding in everyday medical practice and not possible at all in existing data like in available large epidemiological studies like the GNC.

Mature preprocessing pipelines enable image translation in other body regions[8]. For example, in brain MRI, every sample can rigidly be registered to an atlas, and the non-brain tissue is removed. However, in the spine, where vertebrae may be moving between acquisitions, such a simple, rigid preprocessing is impossible. Additionally, the mapping of intensities from the MR to the CT domain is highly dependent on the anatomy: e.g. fat and water would have similar signals in T2w MRI but have substantially different density values in CT, despite being in close anatomical location with a high inter-subject variability. Consequently, a network cannot learn the relationship between anatomy and intensity translation based on unpaired images: The tested unpaired method CUT[7] would require additional constraints to learn an anatomically correct translation. SynDiff[17] has an unpaired CycleGAN[6] in its formulation and worked on paired datasets similar to paired methods. Still, it could not correctly translate the posterior structures on unmatched data. We demonstrated that our rigid registration is a required preprocessing for a correct translation, even for SynDiff, and we believe that better processing, such as deformable registration, can lead to better results. However, to account for inter-vertebra movement between two acquisitions due to different patient lying positions between CT and MR acquisitions would require whole vertebral segmentation. Other papers combat this issue by using axial slices, which only need a local vertebra registration [10–12] or only focusing on the lumbar spine (5–9), where acquisitions can be performed in a more standardized patient positioning than the cervical spine. Oulbacha and Kadourys's et al.[38] also use sagittal slices like our study. However, they face similar challenges with incorrectly translating posterior structures, as observed in their figures. To address these issues, we employed dedicated preprocessing techniques and transitioned to a 3D approach.

## Limitation

Our pipeline enables us to generate segmentations that are available in other modalities. This method cannot produce segmentations of structures that are not segmented but visible in the input domain. We observed weaknesses in translating neck and thoracal regions when using external images, especially for the 2D methods. The posterior elements in the thoracic region were still the most difficult and the segmentation and the translation showed more errors compared to other regions. Classifier-free guidance showed substantial improvement in language based DDIM generation [30], and had a visible impact in 2D translation on an out-of-training distribution like the GNC images. Still, the difference in image quality and the Dice scores are too small to measure. Therefore, we excluded classifier-free guidance[30] from our analysis, as the effect was too small to be investigated in available test sets. The same is true for testing a different number of time steps and the determinism parameter eta. We go in more detail about these inference parameters in the supplemental materials.

## Conclusion

We were able to show that image segmentations can be generated in a novel target domain without manual annotations if segmentations exist for another image domain and paired data for both domains can be obtained. For the spine, we identified minimum registration requirements for paired image-to-image translations. With this approach, SynDiff, Pix2Pix, and DDIM enabled translation of 2D images resulting in similarly good downstream segmentations. We introduced a 3D variant of conditional diffusion for image-to-image translation that improved the segmentation of posterior spinal elements compared to 2D translation. The synthesized segmentations represent a novel ground truth for MRI-based spine segmentations that are prerequisites for spine studies involving large cohorts.

# Supplementary Material

## Paired Image-2-Image

Generative adversarial networks were once the most popular image generation networks until denoising diffusion emerged. A GAN consists of a generator and a discriminator. The generator creates a new image, while the discriminator tries to distinguish between real and synthetic images. This approach leads to sharper images compared to earlier regressing networks like autoencoders, which only minimized a loss function between the generated and predicted image.

Paired image-to-image translation involves having data pairs representing views in distribution A and B. Pix2Pix[4] utilizes image pairs for a direct absolute difference loss, and to prevent blurry outputs, it adds a GAN-Loss to the optimization. However, acquiring paired data can be challenging, and despite Pix2Pix's relatively good performance for its age, there are few fundamentally new models. In GANs, balancing the predictive power of the generator and discriminator is often difficult. An imbalance can cause training to fail, and some GANs may suffer from mode collapse, where only a single image is generated repeatedly. A newer method called denoising diffusion overcomes this issue. For image-two-image translation, we condition our DDIM[3] by concatenating the image pair to the noised input[5], enabling us to learn paired image-to-image translation.

## Unpaired Image-2-Image

Producing unpaired image-to-image datasets is easier but comes with higher unreliability. Most variants use a Cycle-Consistency-Loss[6] to ensure related output images. This loss involves transferring an image to another domain and back, measuring the difference between the original and recreated images. However, this approach falls short when dealing with very imbalanced datasets, as the model may attempt to match the distributions of both domains, even if they are different. Consequently, issues like changing sizes, forgotten elements, or hallucinations may occur. We tested on a model called SynDiff[7], which is similar to CycleGAN. SynDiff includes a CycleGAN that generates image pairs for a DDPM[2]. The DDPM operates in image mode with a fixed step size of 4. On the other hand, CUT[8] stands out as it does not utilize Cycle-Consistency-Loss, but instead, it employs a contrastive loss. In certain layers, there is an enforcement that patches in the same region should be similar across different layers, whereas patches between different regions should be dissimilar.

## Denoising Diffusion

Denoising diffusion is a generative deep learning model. The model is forced to predict a Gaussian noise in an image where the noise has varying strengths. Gaussian noise is purely random, and the model has no choice but to learn how images of the dataset look like. Giving a random noise without an image causes the model to introduce image features into the noise. The noise strengths are defined in t timesteps. Where 0 is the image without noise; in time step t, the image is fully replaced by noise. We always used t=1000 for all experiments. From one step i-1 to i a Gaussian noise of $q(x_i| x_{i-1}) = \mathcal{N}(x_i, \sqrt{1-\beta_i}x_{i-1}, \beta_i I)$ is added where β controls the strength of the noise. Starting at 0 for the noise-free image (i=0) and 1 for the completely noised image (i = t). We use a quadratic cosine curve for beta as in Nichol et al [1]. We optimize the model during training to predict the input noise $\epsilon$ or image $x_0$. As loss, we use the absolute difference loss. We can compute for any timestep a noised image with the "forward formula" $x_i = \sqrt{\overline{\alpha}_i}x_0 + \sqrt{1-\overline{\alpha}_i}\varepsilon$ where ε is a random normal distribution and $\alpha_i = 1 - \beta_i$; $\overline{\alpha}_i = \prod_{j=0}^{i}\alpha_j$. During inference, we iterate over the time steps. The model predicts either the noise $\hat{\epsilon}$ or the final image $\hat{x}_0$. We can compute the other by putting $x_i$ and $\hat{\epsilon}$ or $\hat{x}_0$ into the forward formula and solve for the missing value, like in the case of noise prediction: $\hat{x}_0 = \frac{1}{\sqrt{\overline{\alpha}_i}}(x_i - \sqrt{1-\overline{\alpha}_i}\hat{\epsilon})$. With $\hat{x}_0$ and $x_i$ we can compute the next $x_j$ for the time step j. The denoising diffusion probabilistic model (DDPM) [2] iterates over every step from t to 0. The predicted image $\hat{x}_0$ is mixed with $x_i$, and a new noise with an updated variance σ must be applied. $x_{i-1} = \frac{1}{1-\bar{\alpha}_i}\left(\sqrt{\alpha_i}(1-\bar{\alpha}_{i-1})x_i + \sqrt{\bar{\alpha}_{i-1}}\beta_i\hat{x}_0\right) + \sigma\varepsilon$ where $\sigma = \sqrt{\beta_i(1-\bar{\alpha}_{i-1})/(1-\bar{\alpha}_i)}$. DDIM [3] reduces inference time by skipping an arbitrary amount of time steps. It also introduces a parameter η. For η = 1, we use a random noise ε in each step, while for η = 0, we reuse the predicted noise ε̂. The forward step from step i to j (i>j) is: $\sqrt{\bar{\alpha}_j}\hat{x}_0 + c_1\varepsilon + c_2\hat{\varepsilon}$ where $c_1 = \eta\sqrt{\left(\frac{1-\bar{\alpha}_j}{1-\bar{\alpha}_i}\right)\left(1 - \frac{\bar{\alpha}_i}{\bar{\alpha}_j}\right)}$ and $c_2 = \sqrt{(1-\bar{\alpha}_j) - c_1^2}$. We clamp $\hat{x}_0$ to -1 and 1 because we know we are limited to this range.

## Hyperparameters - SA-UNet

We copied the model parameters from https://github.com/lucidrains/denoising-diffusion-pytorch. We reimplemented the DDIM method to get feature parity to DDPM implementation. The SA-UNet is an often-used architecture for denoising diffusion. [1, 9] It uses self-attention blocks and repeatedly feeds the timestep t as a cosine embedding into the residual blocks. We used a starting channel size of 64. For the details, please refer to our GitHub implementation.

The input is the noised target image and the image of the other domain. The output is the prediction of the noise or denoised image. For the generation, we start with a Gaussian noise and the source image. We use the cosine schedule by Nichol and Dhariwal [1]. We use 1000 sampling steps as a goal for training. For sampling, we use the denoising diffusion implicit formulation. [2, 3] We can reduce this number of steps to 20 without observing a noticeable drop in translation quality. We did not do any further hyperparameter searches for the 3D variants. For the 3D version, we saw that we had to change t to at least 25.

## Hyperparameters – Others

We needed a hyperparameter search for the reference implementation of ResNet and UNet, unlike the SA-UNet, which only required finding a suitable learning rate. For Pix2Pix, we found that a basic UNet works best from the reference implementation. We used five down and up blocks. The batch size was 64 and the dropout was 20 %. For CUT, the optimal hyperparameter was especially difficult. We found that a ResNet with only one down/up convolution works best. We used a batch size of 32 and 8 residual Blocks. We turned off the dropout. The contrastive loss was computed on the 0,4,8,12,16 layers of the reference implementation. [8] The SA-UNet had its contrastive loss computed after every down-convolution and the first up-convolution. The rest of the model is equal to the diffusion implementation. The parameters were fixed on the original T1 dataset. A change in data causes notable differences in performance for CUT and every dataset would have required its own hyperparameter tuning. We had to freeze the parameters for a fair comparison of the model and not to be influenced by our ability to find hyperparameters in an unstable environment.

We used the same discriminator as the reference implementations with three down convolution blocks. [4] The number of channels was the same as the generator network.

## Ablation DDIM Inference

Denoising diffusion was trained to predict a noise at 1000 different strengths. The original denoising diffusion probabilistic model (DDPM) [2] iterates through all noise strengths from the strongest to the weakest noise. DDIM[3] is a reformulation of DDPM and skips an arbitrary number of intermediate noise levels on the same trained network. This reduces the required iteration from 1000 to t. We used t=20 in this paper in 2D and t=25 for 3D. With the reformulated forward process DDIM we get better results than the slower DDPM (**T1**=27.18; **T2**=26.87; p<0.001). For 3D translation, we observe that the spine fully forms with 25 or more inference steps. The background and skull bone were not fully formed with t=25 and would require more timesteps. DDIM has an additional feature where instead of pulling a random noise in every step, we can use only a random noise in the initial step and the other is computed from the previous input and the model prediction. This makes the inference deterministic and can enable interpolation of the DDIM output. This behavior is handled by the parameter $\eta$. $\eta=0$ means fully deterministic and $\eta=1$ means that every step receives a fully random noise. A third inference parameter is classifier free-guidance w [10]. If w is not zero, we sample the model in each step twice. One receives the conditional MRI input, and the other receives a black image. The conditioned output is multiplied by w+1 and the unconditioned output is multiplied by -w. Both are added together. The idea is to push the output away from the general bias of the network towards the condition. All three parameters can be used on a DDPM trained network without requiring retraining and are tested on the 2D network exclusively.

We did an ablation on our DDIM. We only changed one parameter and kept the rest fixed on w=0, $\eta$=1 and t=20. We choose two numbers of timesteps for the ablation t=10, t=20, and t=50. The results lead to no conclusion if there is a better t in image quality. We could reduce the t even further than 20 without sacrificing quality. An $\eta$ of zero has a small negative impact (**T1**=27.48; **T2**=26.81 p<0.001) in noise mode and has a positive impact in image mode (**T1**=27.95; **T2**=27.41; p<0.001). We suspect that the parameters t and $\eta$ have a too minor impact and we cannot say in general what value must be set to get an optimal result. The classifier free-guidance has a small impact on the test data and we see no pattern if any w improves the image quality. We turned off the classifier free-guidance for our 3D models.

The inference hyperparameter of DDIM image mode did not impact the segmentation results, while there are noticeable differences for DDIM in noise mode. The DDIM with t=10 (T1=0.81, T2=0.75, MRSSegClg=0.77) had the best scores in image quality but was the lowest performing inference type in the Dice metric (t=10 vs. t=20 p=0.09). The parameter $\eta$ impacted the Dice score in an inconsistent way. We see no correlation between small differences in the quality metrics and Dice scores.

## Computational Efficiency

We observed that the 3D diffusion converged on a single V40 / RTX 3090 after four days. Which is notably less than the 17.5 days reported by Bieder et al. [11]. The 2D diffusion converged after 2-3 days. Our denoising diffusion models were trained on small batch sizes (2 in 3D and 8 in 2D images) and relatively short compared to other studies. The reason is the single channel output and the image condition easing the diffusion training. For medical data, the image condition reduced the training time by at least half compared to without. On multi-channel images, we noticed a color drift in individual channels in the same magnitude as random Gaussian noise when we trained on batches with 8 batches instead of the recommended 2048. The color drift is moving during training but is mostly fixed during inference. This causes the images to be biased like all images are too red or bright. This effect accumulates, meaning reducing the number of steps reduces the impact of this bias.

**Table S. 1:** Image Quality for T1w and T2w to CT Translation for different quality measures and with DDIM inference ablation.

| From T1w | L1↓ | MSE↓ | PSNR↑ | SSIM↑ | VIFp↑ |
|---|---|---|---|---|---|
| **CUT ResNN (unpaired)** | 0.0224 | 0.0050 | 23.50 | 0.835 | 0.295 |
| CUT SA-UNet (unpaired) | 0.0295 | 0.0083 | 21.76 | 0.819 | 0.269 |
| **Pix2Pix UNet** | 0.0143 | 0.0023 | 27.37 | 0.881 | 0.392 |
| Pix2Pix SA-UNet | 0.0135 | ∴0.0020 | 27.82 | 0.883 | 0.394 |
| **SynDiff (unpaired)** | 0.0150 | 0.0024 | 27.01 | 0.865 | 0.373 |
| DDIM noise $\eta=1$, t=10, w=0 | 0.0135 | 0.0021 | 27.64 | 0.877 | 0.395 |
| DDIM noise $\eta=0$, t=20, w=0 | 0.0139 | 0.0022 | 27.48 | 0.875 | 0.388 |
| **DDIM noise $\eta=1$, t=20, w=0** | 0.0136 | 0.0021 | 27.60 | 0.879 | 0.396 |
| DDIM noise $\eta=1$, t=20, w=1 | 0.0140 | 0.0023 | 27.33 | 0.880 | 0.394 |
| DDIM noise $\eta=1$, t=20, w=2 | 0.0144 | 0.0024 | 27.02 | 0.878 | 0.388 |
| DDIM noise $\eta=1$, t=50, w=0 | 0.0139 | 0.0022 | 27.48 | 0.880 | 0.395 |
| DDPM | 0.0146 | 0.0023 | 27.18 | 0.873 | 0.381 |
| DDIM image $\eta=1$, t=10, w=0 | ∴0.0130 | ∴0.0020 | ∴28.00 | 0.887 | 0.411 |
| DDIM image $\eta=0$, t=20, w=0 | ∴0.0130 | ∴0.0020 | 27.95 | ∴0.889 | ∴0.415 |
| **DDIM image $\eta=1$, t=20, w=0** | 0.0131 | ∴0.0020 | 27.89 | 0.887 | 0.411 |
| DDIM image $\eta=1$, t=20, w=1 | 0.0134 | 0.0021 | 27.65 | 0.887 | 0.408 |
| DDIM image $\eta=1$, t=20, w=2 | 0.0145 | 0.0025 | 26.88 | 0.885 | 0.398 |
| DDIM image $\eta$=1, t=50, w=0 | 0.0136 | 0.0022 | 27.57 | 0.885 | 0.407 |

| From T2w | L1↓ | MSE ↓ | PSNR↑ | SSIM↑ | VIFp ↑ |
|---|---|---|---|---|---|
| **CUT ResNN (unpaired)** | 0.0213 | 0.0046 | 23.72 | 0.848 | 0.312 |
| CUT SA-UNet (unpaired) | 0.0215 | 0.0046 | 23.75 | 0.850 | 0.311 |
| **Pix2Pix UNet** | 0.0142 | 0.0023 | 26.95 | 0.895 | 0.392 |
| Pix2Pix SA-UNet | 0.0142 | 0.0023 | 26.87 | 0.890 | 0.384 |
| SynDiff (unpaired) | 0.0140 | 0.0022 | 27.12 | 0.885 | 0.385 |
| DDIM noise $\eta$=1, t=10, w=0 | 0.0145 | 0.0026 | 26.49 | 0.891 | 0.387 |
| DDIM noise $\eta$=0, t=20, w=0 | 0.0142 | 0.0024 | 26.81 | 0.888 | 0.382 |
| **DDIM noise $\eta$=1, t=20, w=0** | 0.0139 | 0.0023 | 26.92 | 0.894 | 0.391 |
| DDIM noise $\eta$=1, t=20, w=1 | 0.0140 | 0.0023 | 26.81 | 0.894 | 0.386 |
| DDIM noise $\eta$=1, t=20, w=2 | 0.0144 | 0.0024 | 26.59 | 0.891 | 0.379 |
| DDIM noise $\eta$=1, t=50, w=0 | 0.0138 | 0.0022 | 26.99 | 0.894 | 0.389 |
| DDPM | 0.0141 | 0.0023 | 26.87 | 0.890 | 0.381 |
| DDIM image $\eta$=1, t=10, w=0 | 0.0131 | ∴**0.0020** | 27.39 | 0.898 | 0.401 |
| DDIM image $\eta$=0, t=20, w=0 | ∴**0.0130** | ∴**0.0020** | ∴**27.41** | ∴**0.900** | ∴**0.404** |
| **DDIM image $\eta$=1, t=20, w=0** | 0.0131 | 0.0021 | 27.36 | 0.898 | 0.401 |
| DDIM image $\eta$=1, t=20, w=1 | 0.0133 | 0.0021 | 27.27 | 0.897 | 0.398 |
| DDIM image $\eta$=1, t=20, w=2 | 0.0139 | 0.0023 | 26.89 | 0.894 | 0.387 |
| DDIM image $\eta$=1, t=50, w=0 | 0.0132 | 0.0021 | 27.31 | 0.898 | 0.401 |
| | | | | | |
| Pix2Pix 3D | 0.0188 | 0.0039 | 26,38 | 0.889 | 0.428 |
| DDIM 3D noise $\eta$=1, t=25 | 0.0194 | 0.0041 | 26,22 | 0.894 | 0.444 |
| DDIM 3D image $\eta$=1, t=25 | 0.0189 | 0.0040 | 26,22 | 0.892 | 0.434 |

Note. — Arrows indicate if smaller or bigger is better. As a visual aid, we marked the best values with ∴. We marked multiple values if they were below the rounding threshold. MSE = mean squared error, PSNR = peak signal-to-noise ratio, SSIM = structural similarity index metric, VIFp = visual information fidelity, DDIM = denoising diffusion implicit model, DDPM = denoising diffusion probabilistic model, CUT = contrastive unpaired translation, SA-UNet = self-attention U-network

**Table S. 2:** Average Dice score↑ per Volume and per Vertebra on the T1w, T2w, and the MRSSegClg. MRSSegClg (ours) is a split where we improved the segmentation to better align the segmentation with the actual bone structure. This table includes the ablation of DDIM inference.

| Dataset | per vol. T1w | per vert. T1w | per vol. T2w | per vert. T2w | per vol. MRSSegClg | per vert. MRSSegClg | per vol. MRSSegClg (our) | per vert. MRSSegClg (our) |
|---|---|---|---|---|---|---|---|---|
| **CUT ResNN (unpaired)** | 0.30 | 0.28 | 0.49 | 0.46 | 0.54 | 0.49 | 0.54 | 0.50 |
| CUT SA-UNet (unpaired) | 0.09 | 0.08 | 0.26 | 0.23 | 0.02 | 0.01 | 0.03 | 0.02 |
| **Pix2Pix UNet** | 0.79 | 0.80 | 0.73 | 0.69 | 0.75 | 0.74 | 0.76 | 0.76 |
| Pix2Pix SA-UNet | ∴**0.82** | 0.82 | 0.75 | 0.72 | ∴**0.77** | ∴**0.76** | 0.77 | 0.77 |
| **SynDiff (unpaired)** | 0.80 | 0.81 | ∴**0.77** | ∴**0.74** | ∴**0.77** | ∴**0.76** | 0.77 | 0.76 |
| DDIM noise $\eta=1$, t=10, w=0 | 0.76 | 0.77 | 0.65 | 0.61 | 0.74 | 0.72 | 0.77 | 0.78 |
| DDIM noise $\eta=0$, t=20, w=0 | 0.78 | 0.78 | 0.70 | 0.67 | 0.75 | 0.73 | 0.77 | 0.77 |
| **DDIM noise $\eta=1$, t=20, w=0** | 0.78 | 0.77 | 0.72 | 0.69 | 0.75 | 0.73 | 0.77 | 0.78 |
| DDIM noise $\eta=1$, t=20, w=1 | 0.80 | 0.80 | 0.74 | 0.71 | 0.76 | 0.74 | 0.77 | 0.77 |
| DDIM noise $\eta=1$, t=20, w=2 | 0.80 | 0.81 | 0.74 | 0.71 | 0.76 | 0.74 | 0.76 | 0.76 |
| DDIM noise $\eta=1$, t=50, w=0 | 0.81 | 0.82 | 0.73 | 0.70 | 0.76 | 0.74 | 0.77 | 0.78 |
| DDIM image $\eta=1$, t=10, w=0 | 0.81 | 0.82 | 0.75 | 0.72 | ∴**0.77** | ∴**0.76** | ∴**0.78** | ∴**0.79** |
| DDIM image $\eta=0$, t=20, w=0 | ∴**0.82** | 0.82 | 0.75 | 0.72 | ∴**0.77** | ∴**0.76** | ∴**0.78** | 0.78 |
| **DDIM image $\eta=1$, t=20, w=0** | ∴**0.82** | ∴**0.83** | 0.75 | 0.72 | ∴**0.77** | ∴**0.76** | ∴**0.78** | 0.78 |
| DDIM image $\eta=1$, t=20, w=1 | ∴**0.82** | ∴**0.83** | 0.75 | 0.72 | ∴**0.77** | ∴**0.76** | ∴**0.78** | 0.78 |
| DDIM image $\eta=1$, t=20, w=2 | ∴**0.82** | ∴**0.83** | 0.75 | 0.72 | ∴**0.77** | 0.75 | 0.77 | 0.78 |
| DDIM image $\eta=1$, t=50, w=0 | 0.81 | 0.81 | 0.75 | 0.72 | ∴**0.77** | ∴**0.76** | ∴**0.78** | 0.78 |

Note. — We marked the best values with ∴. vol. = volume, vert. = vertebra, MRSSegClg = MRSpineSeg Challenge, DDIM = denoising diffusion implicit model, DDPM = denoising diffusion probabilistic model, CUT = contrastive unpaired translation, SA-UNet = self-attention U-network